\documentclass[RNAAS]{aastex62}

\begin{document}

\title{What Does ``Metallicity" Mean When Interpreting Spectra of Exoplanetary Atmospheres?}

\author{Kevin Heng}
\affiliation{University of Bern, Center for Space and Habitability, Gesellschaftsstrasse 6, CH-3012, Bern, Switzerland}
\email{kevin.heng@csh.unibe.ch}


\section{} 

One of the desired outcomes of studying exoplanetary atmospheres is to set decisive constraints on exoplanet formation theories (e.g., \citealt{madhu14}).  Specifically, astronomers often speak of the ``metallicity" in broad terms.  When modeling the bulk metallicity, workers refer to the elemental iron abundance (Fe/H).  For example, \cite{mf11} and \cite{thorn16} set constraints on the bulk metallicity of non-inflated gas-giant exoplanets by matching predictions from evolutionary models with the measured masses and radii.  By contrast, when exo-atmospheric astronomers speak of the ``metallicity" derived from analysing low-resolution \textit{Hubble} and \textit{Spitzer} spectrophotometry, they are referring to the elemental abundances of oxygen (O/H), carbon (C/H) and nitrogen (N/H)---in decreasing order of importance, since spectra from the \textit{Hubble} Wide Field Camera 3 (WFC3) are primarily sensitive to water (H$_2$O) with secondary contributions from hydrogen cyanide (HCN) and ammonia (NH$_3$), while \textit{Spitzer} photometry is sensitive to methane (CH$_4$) and carbon monoxide (CO).  From retrieving for the water abundances, workers such as \cite{kreidberg14}, \cite{wakeford17,wakeford18}, \cite{arc18} and \cite{mansfield18} have published figures that show the ``metallicity" (in units of ``solar") versus the masses of the exoplanets with entries from the Solar System ice and gas giants overplotted.  (\citealt{nikolov18} published a similar figure using retrieved sodium abundances.)  Even if we assume that bulk and atmospheric elemental abundances are equal, the motivation behind this research note is to demonstrate that the conversion between the water abundance and O/H is not straightforward.

To demonstrate this claim, I use a simple and highly reproducible set of models that include only carbon (C), hydrogen (H) and oxygen (O).  I further assume H$_2$-dominated atmospheres, chemical equilibrium and pure-gas chemistry with the reasoning that the presence of nitrogen, photochemistry, atmospheric mixing and condensation will complicate the issue and not alleviate it.  The water volume mixing ratio (relative abundance by number) is thus the number density of water divided by the number density of H$_2$ (and not H).  The analytical models are taken from \cite{hl16}; their accuracy has been validated by comparison to numerical calculations using the Gibbs free energy minimisation method \citep{ht16}.  To mimic the effects of vertical atmospheric mixing, I assume a range of pressures ranging from 10 mbar to 10 bar.  This is akin to water at chemical equilibrium being mixed up from deeper regions of the atmosphere, a phenomenon known as ``quenching" (e.g., \citealt{tsai17}).  Once mixed up to higher altitudes (lower pressures), the assumption is that the dynamical timescale is much shorter than the chemical timescale, which allows the water abundance to be frozen at its quenched value. 

\begin{figure}[h!]
\begin{center}
\includegraphics[scale=0.5,angle=0]{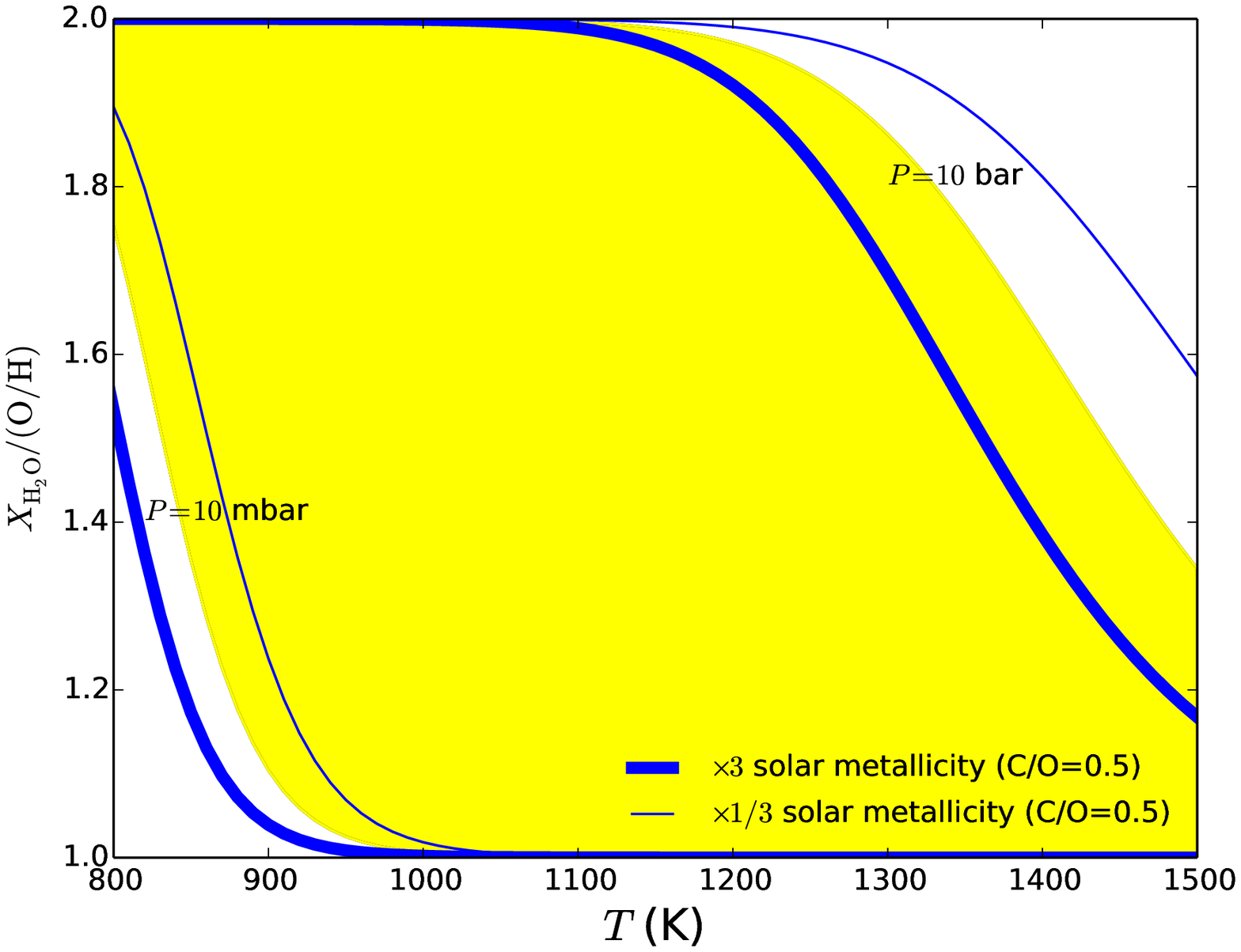}
\includegraphics[scale=0.5,angle=0]{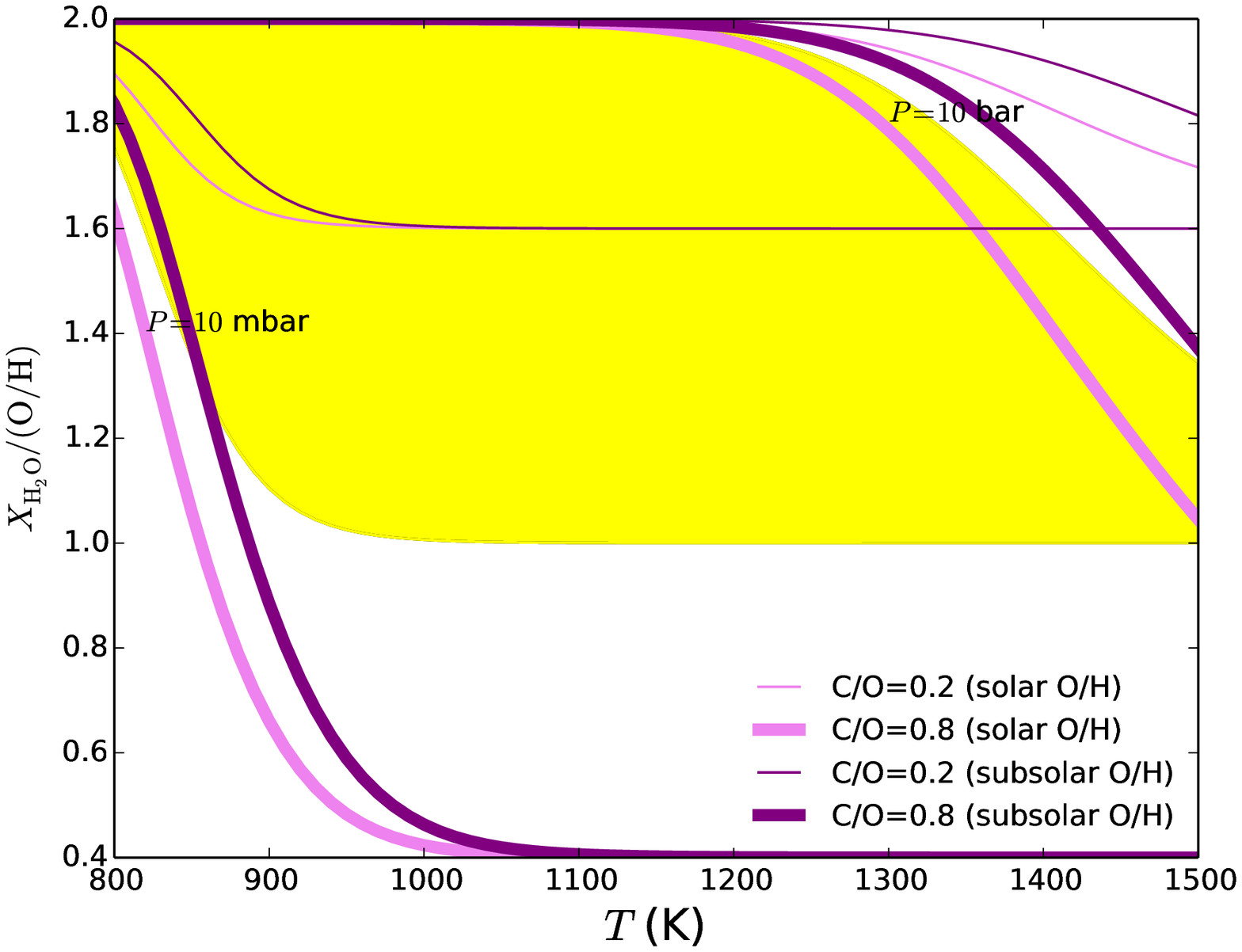}
\caption{Ratio of water volume mixing ratio (relative abundance by number) to the elemental abundance of oxygen (O/H) versus temperature.  The shaded area corresponds to solar O/H ($6\times 10^{-4}$) and $\mbox{C/O}=0.5$, and is bounded by two values of the pressure (10 mbar and 10 bar).  Top panel: C/O is kept at the solar value (0.5), but both C/H and O/H are scaled by the same factor.  Bottom panel: O/H is kept at either the solar value or a subsolar value (1/3 solar), but C/O is allowed to vary (0.2 and 0.8).  Each pair of curves is computed for two pressures: 10 mbar and 10 bar.  It is apparent that the conversion factor between the water volume mixing ratio and O/H is not always unity.}
\label{fig:only}
\end{center}
\end{figure}

It remains unclear what the ``metallicity" means in broad terms; I will work specifically with C/H and O/H.  By ``solar metallicity", I mean that $\mbox{O/H}=6 \times 10^{-4}$ and $\mbox{C/H}=3 \times 10^{-4}$ (such that ``solar C/O" is 0.5).  Even with these definitions, Figure \ref{fig:only} shows that \textit{``water at solar metallicity" is a temperature- and pressure-dependent statement}, because there is no single value of the water volume mixing ratio corresponding to $\mbox{O/H}=6 \times 10^{-4}$ and $\mbox{C/H}=3 \times 10^{-4}$.  By ``$\times 3$ solar metallicity" and ``$\times 1/3$ solar metallicity", I mean that O/H and C/H are scaled up or down by the same factor of 3.  I also consider subsolar O/H (reduced by a factor of 3 from its solar value for illustration) to mimic the effect of oxygen being sequestered in refractory molecules such as enstatite \citep{woitke18}.  Overall, Figure \ref{fig:only} demonstrates that the conversion factor between the water volume mixing ratio and O/H is not unity and depends on temperature, pressure, O/H, C/H and C/O, even without invoking disequilibrium chemistry (photochemistry and/or atmospheric mixing).  The conversion factor is a factor of several.  It remains unproven that disequilibrium chemistry or condensation will \textit{always} return a conversion factor of unity.  It is thus a fair statement to say that the ``metallicity" estimates inferred from atmospheric retrievals are model-dependent.

\acknowledgments

I acknowledge a series of lively question-and-answer sessions at the Exoplanets II conference, which motivated this research note.

\end{document}